\documentclass[a4paper,11pt]{article}
\usepackage{jheppub} 
\usepackage[T1]{fontenc} 
\usepackage{subfig}
\newcommand{\be}{\begin{equation}}
\newcommand{\ee}{\end{equation}}
\newcommand{\Tre}{T_{\rm re}}

\title{\Large \bf Small Field Polynomial Inflation:\\ Reheating,
  Radiative Stability and Lower Bound}
\author[]{Manuel Drees,}
\author[]{Yong Xu}
\affiliation[]{\it Bethe Center for Theoretical Physics and Physikalisches
	Institut, Universit\"at Bonn,\\Nussallee~12, 53115 Bonn, Germany}

\emailAdd{drees@th.physik.uni-bonn.de}
\emailAdd{yongxu@th.physik.uni-bonn.de}

\abstract{We revisit the renormalizable polynomial inflection point
  model of inflation, focusing on the small field scenario which can
  be treated fully analytically. In particular, the running of the
  spectral index is predicted to be
  $\alpha = -1.43 \times 10^{-3} +5.56 \times 10^{-5} \left(N_{\rm
      CMB}-65 \right)$, which might be tested in future. We also
  analyze reheating through perturbative inflaton decays to either
  fermionic or bosonic final states via a trilinear coupling. The
  lower bound on the reheating temperature from successful Big Bang
  nucleosynthesis gives lower bounds for these couplings; on the other
  hand radiative stability of the inflaton potential leads to upper
  bounds. In combination this leads to a lower bound on the location
  $\phi_0$ of the near inflection point, $\phi_0 > 3 \cdot 10^{-5}$ in
  Planckian units. The Hubble parameter during inflation can be as low
  as $H_{\rm inf} \sim 1$ MeV, or as high as $\sim 10^{10}$ GeV.
  Similarly, the reheating temperature can lie between its lower bound
  of $\sim 4$ MeV and about $4 \cdot 10^8 \ (10^{11})$ GeV for
  fermionic (bosonic) inflaton decays. We finally speculate on the
  ``prehistory'' of the universe in this scenario, which might have
  included an epoch of eternal inflation.}

\begin{document} 
	\begin{flushright}
		August 2021
	\end{flushright}
	\maketitle
	\flushbottom
\section{Introduction and Motivations}

Inflation neatly solves the horizon, flatness and monopole problems of
(old) standard cosmology \cite{Starobinsky:1980te,
  Guth:1980zm,Linde:1981mu, Albrecht:1982wi}. The simplest
inflationary model uses a single elementary scalar ``inflaton'' field
$\phi$ to drive slow--roll (SR) inflation, with a monomial
$\lambda\phi^p$ potential; at sufficiently large field values this
even allows eternal inflation (where the inflaton field undergoes
random walk) \cite{Vilenkin:1983xq,Linde:1986fc}. However recent
Planck 2018 measurements \cite{Akrami:2018odb} have disfavored those
models with $p\geq1$: these potentials are too steep and therefore
predict too large a tensor--to--scalar ratio $r$. Agreement with these
observations can be obtained for smaller values of $p$, which however
are not easy to realize in complete particle physics models. We refer
to ref.~\cite{Martin:2013tda} for a review for inflationary modes.

In this paper, we instead consider the most general renormalizable
single--field model, where the potential is a polynomial of degree
four \cite{Hodges:1989dw, Destri:2007pv, Nakayama:2013jka,
  Aslanyan:2015hmi, Musoke:2017frr}. We will assume that the density
perturbations observed in the CMB and other cosmological probes were
produced when the inflaton field had values not larger than the Planck
scale, so that the energy scale during inflation is far below the
Planck scale; hence insisting on renormalizability seems
reasonable. Since the linear term can be removed via a shift of the
inflaton field and the constant term is at most of the order of
today's cosmological constant, which is essentially zero relative to
the energy scales during inflation, the potential only contains three
terms. It turns out that all three terms are needed in order to
reproduce the measurements by the Planck collaboration. In particular,
the potential is sufficiently flat only if it has a (near) inflection
point where both the first and the second derivative of the potential
are very small. Such an inflection point might arise from radiative
corrections \cite{Stewart:1996ey, Stewart:1997wg, Ballesteros:2015noa,
  Dimopoulos:2017xox, Okada:2016ssd, Okada:2017cvy, Okada:2019bqa,
  Okada:2019yne, Okada:2020cvq, Okada:2015lia}, but here we generate
it already at the tree level. Inflation near an inflection point of
the potential has been discussed previously in a supersymmetric
context, often using non--renormalizable potentials or just analyzing
the motion of the field around the inflection point
\cite{Allahverdi:2006iq, Itzhaki:2007nk, Allahverdi:2007wt,
  Badziak:2008gv, Enqvist:2010vd, Hotchkiss:2011am, Gao:2015yha}. It
should be noted that this model does allow for eternal inflation, at
much larger (trans--Planckian) field values but still sub--Planckian
energy densities. Assuming an early phase of ``eternal'' inflation
alleviates the initial condition problem; in fact, it is not clear
whether one can meaningfully speak of ``initial conditions'' in such a
case \cite{Linde:1986fc}. Eternal inflation also offers the only known
physical mechanism that might allow to sample a ``landscape'', i.e. a
(complicated) potential with a very large number of minima
\cite{Susskind:2003kw}.\footnote{For reviews on eternal inflation, see
  e.g. \cite{Guth:2007ng, Winitzki:2006rn}.}

The goal of this paper is to study the non--supersymmetric small field
polynomial inflation model. We wish to explore the entire allowed
parameter space in a complete model, which also includes a coupling
that allows the inflaton to decay; this is required so that the
universe can reheat at the end of inflation. To this end, we first
analytically calculate the number of $e-$folds and inflationary
predictions (power spectrum, tensor--to--scalar ratio, spectral index
and its running). Once the overall size of the density perturbations
and the spectral index have been fixed, essentially only the location
$\phi_0$ of the near--inflection point remains as free parameter. It
is bounded from below by the requirement that the reheating
temperature is sufficiently high \cite{Kawasaki:2000en,
  Hannestad:2004px}, with inflaton couplings that are sufficiently
small not to disturb the flatness of the potential through radiative
corrections. We find that $\phi_0$ has to be larger than
$3 \cdot 10^{-5}$ (in Planckian units). The resulting
tensor--to--scalar ratio is much too small to be detectable. On the
other hand, the running of the spectral index, which turns out to be
independent of $\phi_0$, just might be detectable in future precision
measurements.  Within the allowed parameter space, the inflationary
scale can be as low as $H_{\rm inf} \sim 1$ MeV; such a low
inflationary energy scale might help to embed the QCD axion as dark
matter with a wider cosmologically allowed window, i.e. larger decay
constant $f_a$ than is usually considered \cite{Guth:2018hsa,
  Ho:2019ayl}, and would greatly alleviate the cosmological moduli
problem \cite{Coughlan:1983ci}. On the other hand, for larger (still
sub--Planckian) values of $\phi_0$ the reheat temperature might exceed
$10^{10}$ GeV, which would allow standard thermal leptogenesis
\cite{Fukugita:1986hr, Davidson:2002qv}; however, this requires an
inflationary Hubble parameter of order $10^9$ GeV.

The remainder of this paper is organized as follows. In
Sec.~\ref{setup} we give a complete analytical description of the
small field polynomial inflection point model. In
Sec.~\ref{predictions} the model parameters and predictions of
cosmological observables are investigated. In Sec.~\ref{reheat} we
calculate the reheating temperature and discuss the corresponding
constrains from BBN; we analyze two scenarios, where the inflaton
dominantly decays into two fermions or two bosons, respectively. In
Sec.~\ref{radative}, the radiative stability of the potential under
one--loop corrections is investigated and the resulting lower bound on
$\phi_0$ is derived. In Sec.~\ref{scale}, we investigate the
inflationary scale and reheating temperature within the parameter
space we have obtained.  In Sec.~\ref{prehis}, we briefly describe a
possible ``prehistory'' of our model, starting from a phase of eternal
inflation.  Finally, in Sec.~\ref{sum} we sum up our findings and end
with some prospects to embed our inflation model into some well
motivated BSM scenarios.

\section{The Setup}
\label{setup}

In this Section we introduce our potential. We show that inflation can
occur at small field values, $\phi \leq 1$ in Planckian units, only in
the presence of a very flat region, which requires that the potential
almost possesses a saddle point. If this is the case, the problem can
be treated fully analytically to excellent approximation.

\subsection{The Potential}

A general renormalizable potential of a single real scalar inflaton
$\phi$ has terms $\propto \phi^n$ with $n \in\{0,1,2,3,4\}$. However,
the linear term can be eliminated by shifting the field, such that the
origin is an extremum of the potential. We also neglect the constant
term, which could produce the cosmological constant, which is tiny
compared to the energy scales of interest here. This leaves us with
the potential
\be \label{inflaton_potential}
V(\phi) = b \phi^2 + c \phi^3 + d \phi^4 \,.
\ee
In order to guarantee the potential to be bounded from below we
require $d>0$. The origin is the absolute minimum of the potential if
$b > 0$. Since the potential is invariant under the simultaneous
transformation $ c \to -c $, $\phi \to -\phi$, we can take $c \le 0$
without loss of generality. We work in Planckian units, where the reduced
Planck mass $M_{\rm P} \simeq 2.4 \cdot 10^{18} \ {\rm GeV} \equiv 1$.

The derivatives of the potential are:
\be \label{eq:derivs}
\begin{split}
	V^{\prime}(\phi) &=  2 b \phi + 3 c \phi^2 + 4d \phi^3\,; \\
	V^{\prime \prime}(\phi) &=  2 b  + 6 c \phi + 12 d \phi^2\,.
\end{split}
\ee
At a true saddle point, $V'(\phi_0) = V''(\phi_0) = 0$. This happens
at
\be \label{phi0}
\phi_0 = - \frac{3c}{8d}\,,
\ee
if the parameters of the potential satisfy the relation
\be \label{correlation}
b = \frac{9 c^2}{32d}\,.
\ee
In general the parameters will not obey eq.(\ref{correlation}). Allowing
the cubic term to deviate from this relation by a factor $1-\beta$,
we can rewrite the potential, still in full generality, as
\be  \label{inflaton_potential2}
\begin{split}
	V(\phi) &=   d \left[\phi^4 +  \frac{c}{d}\left( 1- \beta  \right) \phi^3
	+  \frac{9}{32} \left(\frac{c}{d}\right)^2\phi^2\right] \\
	&= d \left[\phi^4 +  A\left(1 - \beta \right)\phi^3
	+  \frac{9}{32} A^2\phi^2\right]\,,
\end{split}
\ee
where we have introduced the quantity
\be \label{eq:A}
A = - \frac{8}{3} \phi_0\,,
\ee
which controls the location of the (would--be) inflection point.

As noted in the Introduction, for small field values ($\phi \leq 1$)
inflation can occur only if the potential indeed ``almost'' has a saddle
point, i.e. $\beta$ has to be small. This can be seen as follows. As well
known, SR inflation requires the parameters $\epsilon_V = 0.5 (V'/V)^2$
and $\eta_V = V''/V$ to be small \cite{Lyth:2009zz}. For our potential, we find
\be \label{SR1}
\begin{split}
	\epsilon_V &= \frac{8}{\phi^2} f(\phi)\,; \\
	\eta_V &= \frac{12}{\phi^2} g(\phi)\,.
\end{split}
\ee
Here the functions $f$ and $g$ approach $1$ for $\phi \gg |A|$; in the
opposite limit, $\phi \ll |A|$, we have
$f(\phi) \rightarrow 1/4, \, g(\phi) \rightarrow 1/6$. ``Generically''
these functions will therefore be of order unity, or slightly
below. Clearly SR inflation would then require $\phi \gg 1$,
i.e. large field values. Here we are interested in small--field
inflation, $\phi \lesssim 1$. Since
$f \propto \left(V'\right)^2,\ g \propto V''$, $\epsilon_V$ and
$\eta_V$ can evidently only be simultaneously small if for some range
of field values both the first and the second derivative of $V$ are
small; which requires the existence of a near saddle point, i.e. we
need $|\beta| \ll 1$.

This parameter controls the flatness of the potential for
$\phi \sim \phi_0$, i.e. the larger $\beta$ is, the more the potential
around $\phi_0$ deviates from a flat plateau. Note that $\beta < 0$
would lead to a negative slope at $\phi_0$, and hence to a second
minimum at some $\phi > \phi_0$. This would require some finetuning of
initial conditions, since the universe could easily get ``stuck'' in
this second minimum if $\phi$ was initially large. We therefore
require $\beta \geq 0$.

As already noted, the model parameter $A$ determines the position of
the saddle point (or flat region of the potential). Finally, the model
parameter $d$ determines the amplitude of the potential, which can be
constrained by the power spectrum near the plateau.

Although the inflaton potential (\ref{inflaton_potential2}) only contains
the three parameters $d, \, A$ and $\beta$, the predictions for cosmological
observables also depend on the value of the inflaton field $\phi$ at
the time when observable density perturbations were produced. As we will
show now, this four--dimensional parameter space can be explored fully
analytically in the region of interest.

\subsection{Analytical Analysis}
\begin{figure}[!ht]
	\centering
	\includegraphics[width=.4\paperwidth, keepaspectratio]{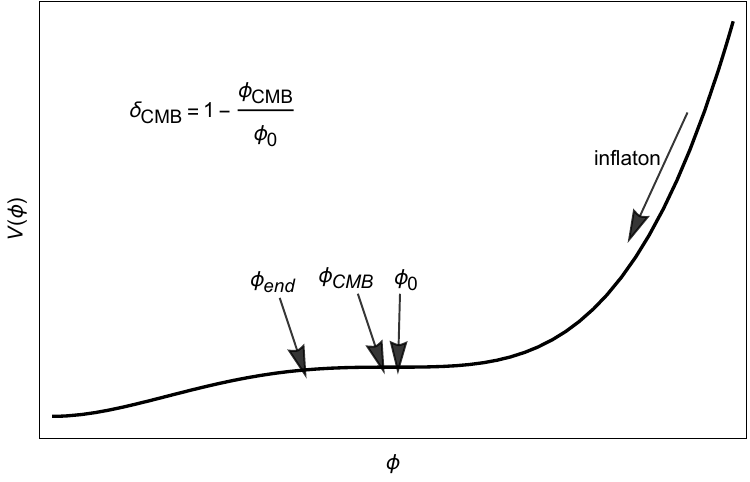}
	\caption {Schematic plot for inflaton potential with an (near)
		inflection-point at $\phi_0$.}
	\label{poential_plot}
\end{figure}

In this paper, we consider $\phi_0 \leq 1$. In this case
$\phi_{\rm CMB} $ (the field value when the ``pivot'' scale
$k_{\star} = 0.05\ \rm{Mpc}^{-1} $ crossed out of the horizon) is very
close to $\phi_0$ (see Fig.~\ref{poential_plot}). We therefore introduce
the field parameter $\delta$:
\be \label{delta}
\phi = \phi_0 (1-\delta)\,,
\ee
so decreasing $\phi$ corresponds to increasing $\delta$. Since both
$\delta$ and $\beta$ are rather small (as we will see,
$\beta\ll \delta \ll 1$), we keep terms up to linear $\beta$ and up to
quadratic in $\delta$ in our analysis, and also drop terms
${\cal O}(\beta \delta)$. 

The following definitions for  SR parameters, number of e-folds and inflationary predictions are based on standard literature, see e.g. Ref.\cite{Lyth:2009zz}. For our model  the SR parameters are given by:
\be \label{srparameters}
\begin{split}
	&\epsilon_V = \frac{1}{2} \left( \frac{V^{\prime}}{V}\right)^2
	\simeq \frac{72 \left(-2\beta (\delta -1) + \delta^2 \right)^2} {\phi_0^2}
	\simeq  \frac{72 \left(2\beta  + \delta^2 \right)^2}{\phi_0^2}\,; \\
	&\eta_V = \frac{V^{\prime \prime}}{V}
	\simeq \frac{12 \left(-4 \beta(\delta -1) + \delta (3\delta -2)\right)}
	{\phi_0^2}
	\simeq \frac{24 \left( 2 \beta - \delta \right)}{\phi_0^2}\,; \\
	&\xi_V^2 = \frac{V^{\prime} V^{\prime \prime \prime}}{V^2}
	\simeq \frac{288 (4 \beta^2 + \beta (2-10\delta) + \delta^2)} {\phi_0^4}
	\simeq \frac{288 (  2\beta  + \delta^2)}{\phi_0^4} \,.
\end{split}
\ee
As already stated, SR requires $\epsilon_V, \, |\eta_V| < 1$. The first
two eqs.(\ref{srparameters}) show that $\epsilon_V \ll |\eta_V|$ in our
case, i.e. the beginning and the end of inflation is determined by
$|\eta_V| = 1$, with
\be \label{end}
\delta_{\rm end} \simeq \phi_0^2 / 24\,.
\ee
The third slow--roll parameter turns out to always be small if
$|\eta_V| < 1$; it affects the running of the spectral index, as we
will see shortly.

The number $N_{\rm CMB}$ of $e-$folds of inflation after the pivot
scale $k_\star = 0.05\ \rm{Mpc}^{-1}$ crossed out of the horizon is
given by:
\be \label{ncmb}
\begin{split}
N_{\rm CMB} &= \int^{\phi_{\rm CMB}}_{\phi_{\rm end}}
\frac{1}{\sqrt{2 \epsilon_{\rm V}}} d\phi \\
& = -\frac{\phi_0^2}{12} \int^{\delta_{\rm CMB}}_{\delta_{\rm end}}
\frac{d \delta }{\left(2\beta  + \delta^2 \right)}\\\
& = -\frac{\phi_0^2}{12 \sqrt{2 \beta}}
\left[ \arctan\left(\frac{\delta_{\rm CMB}}{\sqrt{2 \beta}}\right) -
  \arctan\left(\frac{\delta_{\rm end}}{\sqrt{2 \beta}}\right) \right]\\
&\simeq \frac{\phi_0^2}{12\sqrt{2 \beta}} \left[ \frac{ \pi  }{2}
-\arctan\left(\frac{\delta_{\rm CMB}}{\sqrt{2 \beta}}\right)  \right ]  \,,
\end{split}
\ee
where $\delta_{\rm CMB}$ can be obtained from eq.(\ref{delta}):
\[
\delta_{\rm CMB} = 1 - \frac{\phi_{\rm CMB} }{\phi_0}\,.
\]
In order to resolve the flatness and horizon problems at least $50$
$e-$folds of inflation are needed; in this paper we will take as
typical value $N_{\rm CMB} =65$. Eq.(\ref{ncmb}) then implies
$\sqrt{2\beta} \ll \phi_0^2/12$, i.e.  $\delta_{\rm end}$ of
eq.(\ref{end}) is much larger than $\sqrt{2 \beta}$ so that
$\arctan(\delta_{\rm end}/\sqrt{2\beta}) \simeq \pi/2$.

Eq.(\ref{ncmb}) also shows that $\delta_{\rm CMB}$ cannot be much larger
than $\sqrt{2\beta}$, but it does not exclude the possibility
$\delta_{\rm CMB} \ll \sqrt{2\beta}$. In order to decide this, we look
at the spectral index of the density perturbations :
\be \label{ns}
n_s = 1 - 6\epsilon_V + 2\eta_V
\simeq 1 - \frac{ 48 ( \delta - 2 \beta)}{\phi_0^2}\,.
\ee
Observations imply $n_s < 1$, i.e. we need $\delta_{\rm CMB} > 0$. The
second term in the last line of eq.(\ref{ncmb}) therefore reduces the
number of $e-$folds of inflation. Ignoring this term and requiring
$N_{\rm CMB} > 50$ thus implies $\beta < 3.4 \cdot 10^{-6} \phi_0^4$,
which in turn shows that the term $\propto \beta$ in eq.(\ref{ns}) can
be neglected:
\be \label{deltacmb}
\delta_{\rm CMB} \simeq \left( 1 - n_s \right) \frac {\phi_0^2}{48}\,.
\ee
Eq.(\ref{ncmb}) then requires $\sqrt{2\beta}$ to be of order $\delta_{\rm CMB}$,
so that $\beta \sim {\cal O}(\delta_{\rm CMB}^2) \ll \delta_{\rm CMB}$, as
claimed at the beginning of this Subsection.

During SR inflation, the power spectrum of curvature perturbation can
be approximated by:
\be\label{powerspectrum}
\mathcal{P}_{\zeta} = \frac{V}{24\pi^2\epsilon_V}
\simeq \frac{d \phi_0^6}{5184 \pi^2 (\delta^2 + 2\beta)^2 }\,.
\ee
This is the only observable of interest that depends on the strength
of the quartic coupling $d$.

There are two additional observables, whose values are currently not
so well known but where significant progress is expected in the coming
years. One is the running of the spectral index, which is given by:
\be \label{running}
\alpha = 16\epsilon_V \eta_V - 24 \epsilon_V^2 - 2\xi_V^2
\simeq -\frac{576( 2 \beta +\delta^2)}{\phi_0^4} \,.
\ee
Due to the smallness of $\epsilon_V$, $\alpha$ is dominated by the
contribution $\propto \xi_V^2$, and is negative in our model. The
second observable is the power in gravitational fields produced
during inflation. It is usually described by the
tensor--to--scalar ratio $r$, which is given by:
\be \label{ratio}
r = 16 \epsilon_V
\simeq \frac{1152 \left(2\beta  + \delta^2 \right)^2}{\phi_0^2} \,.
\ee
%

\section{Model Parameters and Inflationary Predictions}
\label{predictions} 

Of course, any potentially realistic model of inflation has to
reproduce known facts. Of particular interest are the
Planck 2018 measurements \cite{Akrami:2018odb} at the pivot scale
$k_\star = 0.05\ \rm{Mpc}^{-1}$:
\be \label{planck2018}
\mathcal{P}_{\zeta}= \left(2.1 \pm 0.1 \right) \times 10^{-9}\,;
\  n_s =  0.9649  \pm 0.0042\,;
\  \alpha = -0.0045\pm0.0067\,;
\   r< 0.061\,.
\ee
We see that two quantities, $\mathcal{P}_\zeta$ and $n_s$, are already
known quite accurately. In addition, we have to satisfy
eq.(\ref{ncmb}) with $N_{\rm CMB} \simeq 65$. Altogether we can thus
essentially fix three of the four free parameters of our model.

We chose to keep $\phi_0$ as a free parameter. The model parameter
$\delta_{\rm CMB}$ is fixed by the spectral index using
eq.(\ref{deltacmb}). Choosing a value of $N_{\rm CMB}$ then fixes
$\beta$ via eq.(\ref{ncmb}). Finally, we use eq.(\ref{powerspectrum})
to fix the quartic coupling $d$.  	

For the central values of $n_s$ and $\mathcal{P}_\zeta$ and our
standard choice $N_{\rm CMB} = 65$ we find in this way:
\be \label{pa1}
\delta_{\rm CMB}  = 7.31 \times 10^{-4} \phi_0^2\,;
\ee
\be \label{pa2}
\beta = 9.73 \times 10^{-7} \phi_0^4\,; 
\ee
\be \label{pa3}
d = 6.61 \times 10^{-16} \phi_0^2\,.
\ee
The scaling with powers of $\phi_0$ can be traced back to eq.(\ref{deltacmb});
the numerical factor in
(\ref{pa1}) corresponds to the result with $n_s = 0.9649$ and
$\phi_0 =1$. Since $\delta_{\rm{CMB}} \propto \phi_0^2$, we see from
(\ref{ncmb}) that $\beta$ should be $\propto \phi_0^4$ in
order to yield a fixed $N_{\rm CMB}$. The numerical pre--factor in
(\ref{pa2}) comes from the numerical factor in (\ref{pa1}) and
$N_{\rm CMB} = 65$. Finally, $d \propto \phi_0^2$ (from
eq.(\ref{powerspectrum})) is required to have a fixed power
$\mathcal{P}_{\zeta} = 2.1 \times 10^{-9}$.

With eqs.(\ref{pa1}), (\ref{pa2}), (\ref{pa3}) and (\ref{ratio}),
one obtains the prediction
\be \label{rvalue}
r = 7.09\times 10^{-9} \phi_0^6 \,.
\ee
For $\phi_0 \leq 1$ this is well below the sensitivity of any currently
conceivable observation. Varying $N_{\rm CMB}$ and $n_s$ over their allowed
ranges does not change this conclusion. On the other hand, eq.(\ref{running})
predicts for the running of spectral index
\be \label{avalue}
\alpha = -1.43 \times 10^{-3}\,.
\ee
This might be within the sensitivity of a combination of future CMB
measurements with greatly improved investigations of structures at
smaller scale, in particular the so--called Lyman$-\alpha$ forest
\cite{Munoz:2016owz}. We note that $\alpha$ is independent of
$\phi_0$, i.e. this is a clear prediction of our model.

Recently there has been quite a bit of interest in production
mechanisms of primordial black holes (PBHs). In principle they can be
produced by the gravitational collapse of domains that have a high
over--density after inflation. However, this requires \cite{Carr:2020xqk} a power of $\mathcal{O}(10^{-2})$. From eq.(\ref{powerspectrum}) we see
that in our model the highest power occurs at $\delta = 0$; however,
eqs.(\ref{pa2}) and (\ref{pa3}) show that this maximal power only
amounts to $\mathcal{O}(10^{-8})$, independent of $\phi_0$. Moreover,
eq.(\ref{powerspectrum}) implies that the power decreases
monotonically as $\delta$ increases, i.e. with decreasing scale, as
also indicated by $n_s < 1$ and $\alpha < 0$.  Hence the current model
does not lead to PBH formation from primordial density fluctuations.

\begin{figure}[ht!] 
\centering
\includegraphics[width=.6\paperwidth, keepaspectratio]{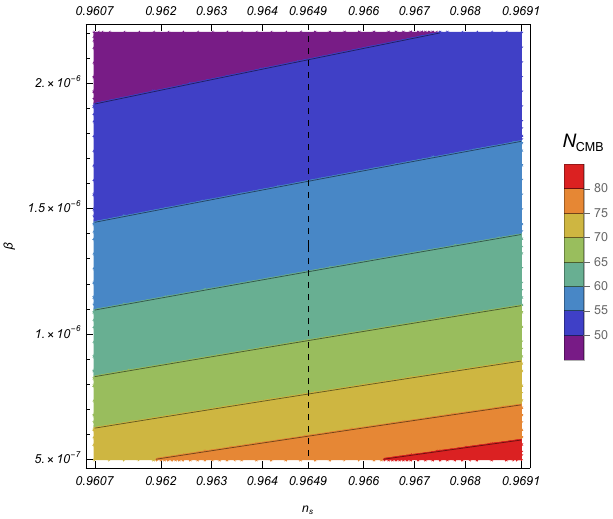}
\caption {$N_{\rm CMB}$ as function of $n_s$ and $\beta$ for
	$\phi_0 =1$; for other values of $\phi_0$, $\beta$ has to be
	rescaled by $\phi_0^4$. The vertical black line denotes the current
	central value of $n_s$, which crosses the contour line with
	$N_{\rm CMB} =65$ for $\beta = 9.73\times 10^{-7}$.}
\label{beta_ns}
\end{figure}

Eqs.(\ref{pa1}) to (\ref{avalue}) hold for the central value of $n_s$
and $N_{\rm CMB} = 65$. Deviations from these values are explored in
Fig.~\ref{beta_ns}. We see that $\beta$ is of order $10^{-6} \phi_0^4$
for the entire allowed parameter space. The results shown in this
figure can again be understood analytically. To that end we first
expand (around the central values)
\be \label{ee}
1 - n_s = 0.0351 (1+\epsilon_n)
\ee
and 
\be \label{eb}
\sqrt{\frac{\beta}{\phi_0^4}} = 9.86 \cdot 10^{-4} (1+\epsilon_b)\,.
\ee
Taylor expanding the $\arctan$ function in eq.(\ref{ncmb}) around the central
value then yields:
\be \label{epsilonb}
\epsilon_b = \frac{65- N_{\rm CMB} }{40.4} - 0.61 \epsilon_n\,.
\ee
Eq.(\ref{epsilonb}) enables us to obtain an analytical expression for
$\beta$ as function of $N_{\rm CMB}$ and $n_s$:
\be \label{beta}
\sqrt{\frac{\beta}{\phi_0^4}}  = 9.86 \times 10^{-4} \left\{
1 + \left[ \frac {65- N_{\rm CMB} } {40.4}
- 0.61 \left(\frac{1-n_s}{0.0351} -1 \right) \right]\right\}\,,
\ee
which agrees very well with the numerical results shown in
Fig.~\ref{beta_ns}.

\begin{figure}[ht!]
	\centering
	\includegraphics[width=.6\paperwidth, keepaspectratio]{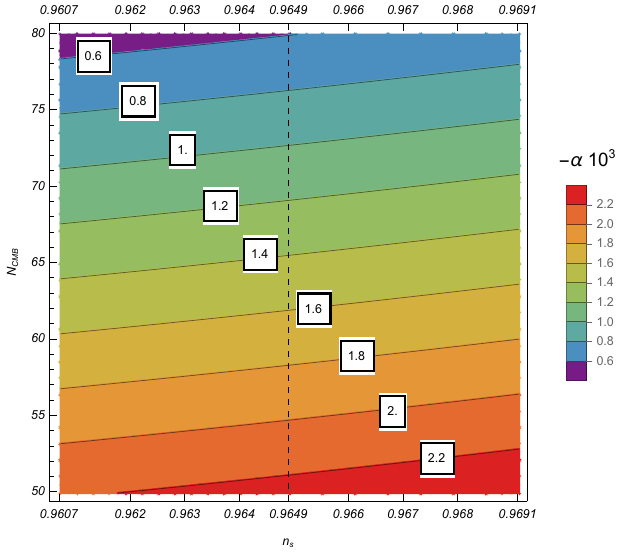}
	\caption {Prediction for the running of the spectral index
		$-\alpha/10^{-3}$ as function of $n_s$ and $N_{\rm CMB}$. Our model
		predicts $\alpha \sim -10^{-3}$ when $n_s$ lies in the vicinity of
		its current central value (vertical black dashed line) and
		$50< N_{\rm CMB} < 80$, which might be testable in future.}
	\label{ncmb_ns}
\end{figure}

As already noted, $r$ remains tiny, of order $10^{-8} \phi_0^6$, over
the entire allowed parameter space. The dependence of the running of
the spectral index $\alpha$ on $n_s$ and $N_{\rm CMB}$ is given by
\be \label{avalue2}
\begin{split}
	\alpha &=  -\frac{576( 2 \beta +\delta^2)}{\phi_0^4}\\
	&= -1.43 \cdot 10^{-3} - 5.56 \cdot 10^{-5} \Big[65- N_{\rm CMB}\Big]
	+ 0.02149 \Big[ 0.9649-n_s\Big] -0.25  \Big[0.9649-n_s\Big]^2\,,\\
\end{split}
\ee
which still does not depend on $\phi_0$; the result is shown in
Fig.~\ref{ncmb_ns}. 

Evidently our model requires a very small but positive value of
$\beta$, see eq.(\ref{pa2}). Eq.(\ref{beta}) shows that $\beta$ varies
approximately linearly when $n_s$ and/or $N_{\rm CMB}$ are varied over
their allowed ranges. In that sense $\beta$, while undoubtedly very
small, is not very finely tuned. On the other hand, setting
$\beta = 0$ does not enhance the symmetry of the potential. This means
that radiative corrections to $\beta$ -- or, more accurately, to the
first and second derivative of the potential at $\phi = \phi_0$ --
need not be proportional to $\beta$.  In order to compute these
corrections, we first have to expand the scope of our model to include
reheating. After inflation the inflaton field has to decay away to
produce relativistic Standard Model particles, i.e. radiation;
otherwise no hot Big Bang will result. This requires some coupling(s)
of the inflaton to lighter particles. These couplings will contribute
to the radiative corrections to the inflaton potential. Before
computing these corrections, we therefore need to discuss reheating.

\section{Reheating}
\label{reheat}
After inflation ends, the inflaton field oscillates around the minimum
of its potential and transfers energy to other degree of
freedoms. This process is usually called reheating.\footnote{For
  reviews on (p)reheating, see e.g. Refs. \cite{Allahverdi:2010xz,
    Amin:2014eta, Lozanov:2019jxc}.} In general it consists of a
non--perturbative ``preheating'' stage followed by the perturbative
decay of the remaining inflaton particles
\cite{Kofman:1997yn}. Finally, the decay products have to thermalize.

In this paper we focus on the simplest possibility, perturbative
decays of the inflaton through trilinear couplings of the inflaton to
lighter particles. Such a coupling is always required, since otherwise
at least some inflaton particles would remain at the end of the
reheating period. At the end of this Section we will argue that
non--perturbative effects cannot significantly deplete the inflaton
number in our model; and in the next Section we will see that all
inflaton couplings have to be so small that inflaton annihilation
reactions are completely negligible.

We compute the reheating temperature in the instantaneous decay
approximation. By setting the energy density in inflaton matter,
$\rho_\phi = m_\phi n_\phi$, equal to the radiation density
$\rho_R = \pi^2 g_* T_{\rm re}^4/30$ at time
$t = 2/(3H) = 1/\Gamma_\phi$, with $H^2 = \rho/3$ as usual in FRW
cosmology, we find (still using Planckian units)
\be \label{tre}
\Tre \simeq 1.41 g_{\star}^{-1/4} \Gamma_\phi^{1/2}\,.
\ee
Here $g_*$ is the number of light degrees of freedom forming the
thermal plasma, and $\Gamma_\phi$ is the perturbative inflaton decay
width. For $\Tre > 1$ GeV, $g_*$ is of order $100$. 

For completeness we allow the inflaton to decay into a Dirac fermion
$\chi$ and/or a scalar $\phi'$. Since $\phi$ is a singlet under the SM
gauge group, $\chi$ would have to be in a vector--like representation
of that group, i.e.  it cannot be an SM fermion. On the other hand,
$\phi'$ might be the Higgs field of the SM. We will treat this as our
standard case, i.e. we will assume that $\phi'$ contains four degrees
of freedom, just like the Dirac fermion $\chi$. The relevant parts of
the Lagrangian are given by
\be \label{lag}
\mathcal{L} = i \bar{\chi} \gamma^\mu\partial_\mu \chi +
(\partial_\mu \phi')^\dagger \partial^\mu \phi'
- m_\chi \bar{\chi}\chi - y  \phi \bar{\chi} \chi
- m_{\phi'}^2 |\phi'|^2 - g \phi |\phi'|^2 + \lambda_{\phi^{\prime}} |\phi'|^4 \,.
\ee
The total decay width of the inflaton is then given by
\be \label{decay_rate}
\Gamma_\phi = \frac{y_{\chi}^2 m_{\phi}}{8\pi}
\left( 1- \frac{4 m_{\chi}^2}{m_{\phi}^2}\right)^{3/2}
+ \frac{g^2}{8\pi m_{\phi}} \sqrt{1- \frac{4m_{\phi^{\prime}}^2} {m_{\phi}^2}}
\simeq  \frac{y^2}{8 \pi} m_{\phi} + \frac{g^2 }{8 \pi m_{\phi}}\,,
\ee
where the inflaton mass is\footnote{This is the mass of inflaton particles
	after inflation, and therefore not directly related to the SR parameter
	$\eta_V$, which is also computed from the second derivative of the
	inflaton potential.}
\be \label{mphi}
m^2_\phi = \left. \frac{\partial^2 V}{\partial \phi^2}\right|_{\phi = 0}
\simeq \frac{9}{16}dA^2 = 4d \phi_0^2\,.
\ee
In the numerical analysis we will assume that one of the two terms
in eq.(\ref{decay_rate}) dominates; the other one may then even vanish.
Moreover, we assume that the mass of $\chi$ or $\phi'$ is much smaller
than $m_\phi$; this minimizes the upper bound on the couplings $y$ and $g$
which we wish to derive.

If fermionic decays dominate, the reheat temperature is given by
\be \label{tref}
\Tre^\chi \simeq 1.41 g_{\star}^{-1/4}
\left(2 \phi_0 \frac{y^2}{8 \pi} \sqrt{d}  \right)^{1/2}\,,
\ee
while for the bosonic decay mode,
\be \label{treb}
\Tre^{\phi'} \simeq 1.41 g_{\star}^{-1/4}
\left(\frac{g^2}{8 \pi \,2 \phi_0  \sqrt{d}} \right)^{1/2}\,.
\ee
Successful BBN requires $\Tre \geq 4$ {\rm MeV}, i.e.
$\Tre \geq 1.67\cdot10^{-21}$ in Planckian units
\cite{Kawasaki:2000en,Hannestad:2004px}. Taking $g_{\star} =10.75$,
as appropriate for a temperature of 4 MeV, and
$d = 6.61 \times 10^{-16} \phi_0^2$ from Eq.(\ref{pa3}), we finally obtain 
lower bounds on the inflaton couplings:
\be \label{bbn_constrain}
y \phi_0  \geq 4.7 \times 10^{-17} \,,
\ee
if fermionic decays dominate, and
\be \label{bbn_constrain2}
\frac{g}{\phi_0} \geq 2.4 \times 10^{-24}\,,
\ee
for bosonic inflaton decays. The scaling with $\phi_0$ can be understood from
the observation that $m_\phi \propto \sqrt{d}\phi_0 \propto \phi_0^2$,
from eqs.(\ref{mphi}) and (\ref{pa3}). Eq.(\ref{tre}) shows that a constant
reheat temperature corresponds to a constant decay width $\Gamma_\phi$.
From eq.(\ref{decay_rate}) this requires constant $y^2 m_\phi$, i.e. constant
$y^2 \phi_0^2$, if fermionic decays dominate, but constant $g^2/m_\phi$,
i.e. constant $g^2/\phi_0^2$ for bosonic inflaton decays; note that $g$
has dimension of mass in natural units, whereas $y$ is dimensionless.

Before concluding this Section we come back to the issue of
non--perturbative inflaton depletion mechanisms. In principle the
couplings in the Lagrangian (\ref{lag}) allow both fermionic and
bosonic preheating. However, due to Pauli blocking, fermionic
preheating is usually very inefficient \cite{Greene:1998nh,
  Greene:2000ew}, i.e. it reduces the initial inflaton energy by less
than $1\%$.

For bosonic inflaton decays, our trilinear $\phi |\phi'|^2$ coupling
leads to a tachyonic instability if $|g \phi| > m^2_{\phi'}$, where
the squared mass of the daughter particle is negative for part of each
oscillation of the inflaton field. As shown in \cite{Dufaux:2006ee, Abolhasani:2009nb} this can build up a sizable $\phi'$ number
density after a very small number of $\phi$ oscillations. However,
even with very small $\phi'$ self--interactions of $\mathcal{O}(10^{-6})$ this only allows to
transfer less than $10\%$ of the initial inflaton energy
\cite{Dufaux:2006ee}. In our default scenario, where $\phi'$ is the SM
Higgs field, $\phi'$ does have a sizable self interaction of the form
$\lambda_{\phi^{\prime}} \phi^{\prime\,4}$. This gives an extra
positive effective squared mass
$\sim \lambda_{\phi^{\prime}} \langle \phi^{\prime 2} \rangle$ (with
$\langle \phi^{\prime 2} \rangle$ denoting the variance) once the
daughter $\phi^{\prime}$ particles are copiously produced. This
back--reaction counteracts the negative tachyonic mass and quickly
terminates preheating, making it even less efficient
\cite{Dufaux:2006ee}; in a lattice simulation \cite{Felder:2000hq} we found that less than $1\%$ of the initial inflaton
energy is depleted through preheating also for these bosonic decay
products. Preheating can thus indeed be neglected in our
model.\footnote{In the literature on preheating some scenarios have
  been suggested which could deplete inflaton energy more efficiently,
  see e.g. instant preheating \cite{Felder:1998vq} or combined
  reheating \cite{Garcia-Bellido:2008ycs, Repond:2016sol,
    Fan:2021otj}. Here one introduces additional perturbative decay
  channels for the produced daughter particles in order to get rid of
  the back reaction problem. However, the decay width of the SM Higgs,
  our default bosonic decay product, is just $4$ MeV, which is $5$ orders of magnitude smaller than the minimal allowed value of
  $m_\phi$ in our model (see below). The preheating time scale, which
  is ${\cal O}(1/m_\phi)$, is thus very much shorter than the $\phi'$
  lifetime, in which case $\phi'$ decays cannot affect the preheating
  dynamics.}

\section{Radiative Corrections and Stability}
\label{radative}

The lower bounds (\ref{bbn_constrain}) and (\ref{bbn_constrain2}) on the
inflaton couplings imply lower bounds on the radiative corrections to
the inflaton potential caused by these couplings. The self--couplings of
the inflaton, described by the potential (\ref{inflaton_potential2}),
also contribute to the radiative corrections. In this Section we
investigate the impact of these corrections in 1--loop order. This will
lead to upper bounds on the couplings; together with the lower bounds derived
in the previous Section this will finally yield a lower bound on the
remaining free parameter $\phi_0$.

The starting point of this analysis is the expression for the
1--loop effective potential, in the formalism of Coleman and Weinberg
(CW) \cite{Coleman:1973jx}:
\be \label{cw}
\Delta V(\phi) = \frac{1}{64 \pi^2} \sum_\psi (-1)^{2 s_\psi} g_\psi
\widetilde{m}_\psi(\phi)^4 \left(
\ln \left(\frac{\widetilde{m}_\psi(\phi)^2}{Q_0^2} \right)
- \frac{3}{2} \right)\,.
\ee
The sum runs over all fields $\psi$ that couple to the inflaton field
$\phi$. $s_\psi$ is the spin of $\psi$; the factor $(-1)^{2 s_\psi}$
therefore implies that bosons (fermions) contribute with positive
(negative) sign to $\Delta V$. $g_\psi$ is the number of degrees of
freedom of the field $\psi$; it includes a spin multiplicity factor
$2 s_\psi + 1$.  Finally, $\widetilde{m}_\psi(\phi)$ is the
$\phi-$dependent mass of $\psi$ (not to be confused with the physical
mass), and $Q_0$ is a renormalization scale.

In our case, up to three fields couple to the inflaton: the inflaton
itself, as well as the fermionic and bosonic decay products $\chi$ and
$\phi'$ introduced in the previous Section. Their field--dependent
masses are given by:
\be \label{masses}
\begin{split}
	\widetilde{m}_\phi^2(\phi) &= 12 d \phi^2 + 6 d A (1-\beta) \phi
	+ \frac{9}{16} d A^2\,;  \\
	\widetilde{m}_\chi^2(\phi) &= \left(m_\chi + y \phi \right)^2\,; \\
	\widetilde{m}^2_{\phi'}(\phi) &= m^2_{\phi'} + g\phi\,.
\end{split}
\ee

We want to make sure that the predictions derived in Sec.~3 are stable
under radiative corrections. To this end we need to investigate the
potential around the point $\phi_0$, where inflation happens. In fact,
the tree--level potential $V_0$ itself is not particularly suppressed
at $\phi = \phi_0$: $V_0(\phi_0) \rightarrow d \phi_0^4/3$ as
$\beta \rightarrow 0$. On the other hand, it is essential that the
first and second derivatives of the potential {\em are} suppressed at
$\phi_0$; this is why $\phi_0$ is a near inflection point. Recall also
that $V_0^\prime$ and $V_0^{\prime\prime}$ directly determine
$N_{\rm CMB}$ and $n_s$, respectively. From
eq.(\ref{inflaton_potential2}) with $A = - 8 \phi_0/3$ we have
\be \label{v0der}
\begin{split}
	V_0^\prime(\phi_0) &= 8 d \beta \phi_0^3\,; \\
	V_0^{\prime\prime}(\phi_0) &= 16 d \beta \phi_0^2\,.
\end{split}
\ee
On the other hand, from eq.(\ref{cw}) the derivatives of the CW correction
to the potential can be written as
\be \label{cwder}
\begin{split}
	\Delta V^\prime = \frac{1}{32 \pi^2} \sum_\psi (-1)^{2 s_\psi} g_\psi
	\widetilde{m}_\psi^2 \widetilde{m}_\psi^{2\prime}
	\left( \ln \left(\frac{\widetilde{m}_\psi^2}{Q_0^2} \right)
	- 1 \right)\,; \\
	\Delta V^{\prime\prime} = \frac{1}{32 \pi^2} \sum_\psi (-1)^{2 s_\psi} g_\psi
	\left\{ \left[ \left(\widetilde{m}_\psi^{2\prime} \right)^2
	+ \widetilde{m}_\psi^2 \widetilde{m}_\psi^{2\prime\prime} \right]
	\ln \left(\frac{\widetilde{m}_\psi^2}{Q_0^2} \right)
	- \widetilde{m}_\psi^2 \widetilde{m}_\psi^{2\prime\prime}  \right\}\,.
\end{split}
\ee
Here $\widetilde{m}_\psi^{2\prime}$ and $\widetilde{m}_\psi^{2\prime\prime}$
are the first and second derivatives of $\widetilde{m}_\psi^2$ with
respect to $\phi$.

The loop corrections are minimized if the bare masses $m_\chi$ and
$m_{\phi'}$ vanish. Recall also that these masses must be below half
the physical inflaton mass; using eqs.(\ref{mphi}) and (\ref{pa3})
this implies
$m_\chi, \ m_{\phi'} < \sqrt{d} \phi_0 = 2.6 \cdot 10^{-8} \phi_0^2$,
which is already quite small. In the subsequent analysis we will
therefore assume $m_\chi \ll y \phi_0$ and $m^2_{\phi'} \ll g \phi_0$,
so that the bare mass terms can be neglected. Moreover, we set
$Q_0 = \phi_0$, since this is the field value we are interested in;
this means that the Lagrangian parameters $y$ and $g$ should be
interpreted as running couplings, taken at scale $Q_0$. The
derivatives of the correction to the potential at $\phi = \phi_0$ are
then given by:
\be \label{cwder1}
\begin{split}
	\Delta V^{\prime} (\phi_0) &= \frac{\phi_0^3}{4\pi^2}
	\left[y^4 - 16 d^2 \beta - y^4 \ln(y^2)
	+ 16 d^2 \beta \ln(16d \beta) \right] 
	+ \frac{ g^2 \phi_0} {8 \pi^2} \left[ \ln\left(\frac{g}{\phi_0}\right) - 1
	\right] \,; \\
	\Delta V^{\prime \prime} (\phi_0) &= \frac{\phi_0^2} {4\pi^2}
	\left[ y^4 - 3y^4 \ln (y^2) + 8d^2 \ln(16 d \beta)  \right]
	+ \frac{ g^2} {8 \pi^2} \ln\left(\frac{g }{\phi_0}\right) \,.
\end{split}
\ee
In the first eq.(\ref{cwder1}) we have ignored terms of order $d^2\beta^2$.
We see that all corrections from the inflaton self--coupling $d$ are
proportional to $\beta$, which means that these terms are automatically
smaller than the tree--level result given in the first eq.(\ref{v0der}).
In the second eq.(\ref{cwder1}) we neglected also terms linear in $\beta$.
We see that nevertheless a finite one--loop correction $\propto d^2$
remains. 

In order to ensure stability of our inflationary model against
radiative corrections, we will require that the terms
$\propto d^2, \, \propto y^4$ and $\propto g^2$ are separately smaller
than the tree--level results of eqs.(\ref{v0der}). We just saw that in
case of $d^2$ only the correction to the second derivative of the
potential can be dangerous. Demanding that it is smaller in magnitude
than the tree--level result leads to the constraint
\be \label{dbound}
\Bigl\lvert \frac{d^2  \ln(16 d \beta)}{\pi^2} \Bigr\rvert < 8 d \beta\,.
\ee
Using the numerical values from eqs.(\ref{pa2}) and (\ref{pa3}) this
implies
$$|\ln(10^{-20} \phi_0^6)| < 1.16 \cdot 10^{11} \phi_0^2\,,$$
which in turn implies
\be \label{phi0d}
\phi_0 > 3 \cdot 10^{-5} \,.
\ee

The strongest upper bound on the Yukawa coupling also comes from the
second derivative of the potential:
\be \label{ybound}
\Bigl\lvert \frac{y^4 - 3y^4 \ln (y^2) } {4\pi^2} \Bigr\rvert < 16 d \beta\,.
\ee
In order to turn this into a lower bound on $\phi_0$, we again use
eqs.(\ref{pa2}) and (\ref{pa3}) for the right--hand side, and insert
the lower limit (\ref{bbn_constrain}) from reheating for $y$; this gives
\be \label{phi0y}
\phi_0 > 3.4 \cdot 10^{-5} \,,
\ee
which is slightly stronger than the bound (\ref{phi0d}). 

On the other hand, the strongest bound on the coupling $g$ originates
from the first derivative of the potential; it reads
\be \label{gbound}
\frac {g^2} {8 \pi^2} \Bigl\lvert \ln\left(\frac{g}{\phi_0}\right) - 1
\Bigr\lvert < 8d\beta \phi_0^2\,.
\ee
Replacing $g$ by its lower bound (\ref{bbn_constrain2}) then implies
\be \label{phi0g}
\phi_0 > 3.1 \cdot 10^{-5}\,,
\ee
very close to the bound (\ref{phi0d}) which is independent of reheating.

\begin{figure}[!ht]
	\centering
	\includegraphics[width=.35\paperwidth, keepaspectratio]{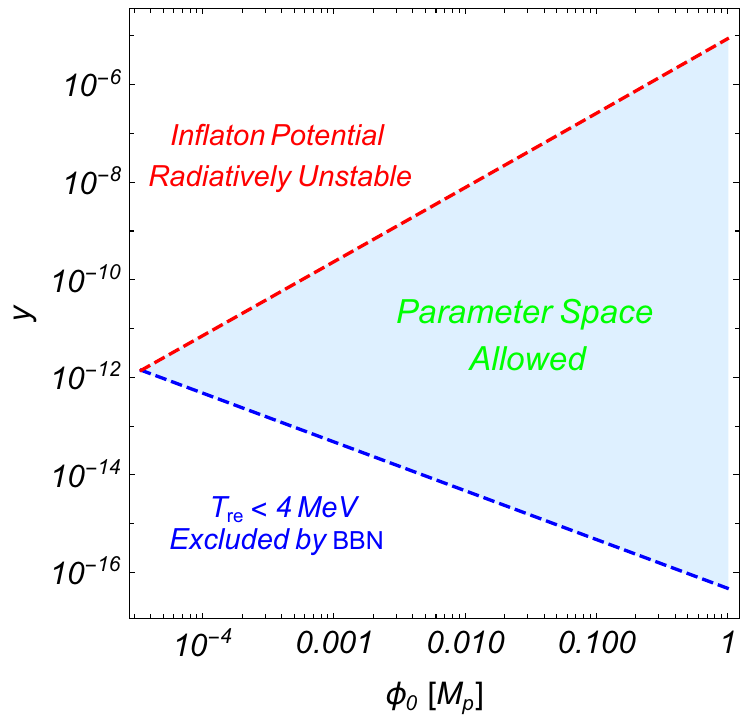}
	\includegraphics[width=.35\paperwidth, keepaspectratio]{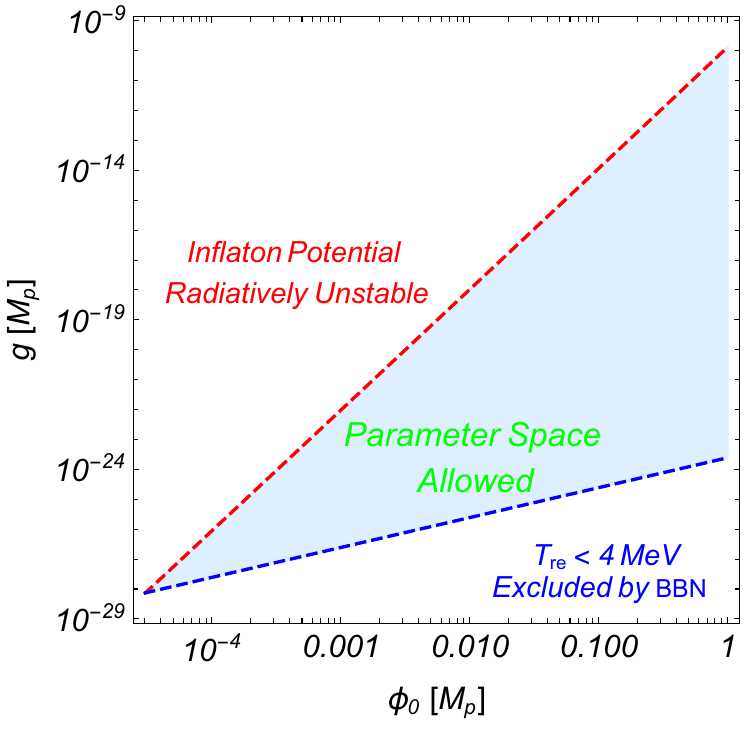}
	\caption {The light blue region is the allowed parameter space,
		yielding $\Tre\geq 4$ MeV while keeping the inflaton potential
		stable against radiative corrections. The left (right) frame is for
		fermionic (bosonic) inflaton decays.}
	\label{ps}
\end{figure}

The constraints on the parameter space spanned by $\phi_0$ and the
coupling that is responsible for reheating are shown in Fig.~\ref{ps};
the left and right frames are for fermionic and bosonic inflaton decays,
respectively. The allowed parameter space, shown in blue, ends at the
values of $\phi_0$ given by the bounds (\ref{phi0y}) and (\ref{phi0g}).
Evidently the allowed range of couplings opens up when $\phi_0$ increases;
for the maximal value we consider, $\phi_0 = 1$, it ranges over $11$
orders of magnitude for $y$, and $13$ orders of magnitude for $g$.
Nevertheless, even for $\phi_0 = 1$ the maximal allowed value of the
Yukawa coupling is about $10^{-5}$, which is only slightly larger than
the Yukawa coupling of the electron in the SM.

Recall that we assumed that four (bosonic or fermionic) degrees of
freedom couple to the inflaton, i.e.  $g_{\phi'}=g_\chi=4$. In case of
bosonic decays, both the lower bound on $g^2$ from reheating and the
upper bound from radiative stability scale like $1/g_{\phi'}$,
i.e. the resulting lower bound (\ref{phi0g}) does not depend on
$g_{\phi'}$. On the other hand, for fermionic decays the lower bound
on $y^2$ scales like $1/g_\chi$ while the upper bound scales like
$1/\sqrt{g_\chi}$; the bound (\ref{phi0y}) therefore roughly scales
like $g_\chi^{-1/10}$. However, it is in any case already quite close
to the bound (\ref{phi0d}) which is independent of reheating.

We ignored a possible quartic coupling $\lambda \phi^2 |\phi'|^2$ in
our discussion of (p)reheating. Such a coupling would also contribute
to the CW corrections to the potential. Demanding that this
contribution to the first derivative of the inflaton potential at
$\phi_0$ does not exceed the tree--level value gives the quite
stringent upper bound
$\lambda \sqrt{|\ln(\lambda)|} < 4.5 \cdot 10^{-10} \phi_0^3$. The
largest quartic coupling $\lambda$ allowed by this bound is of
$\mathcal{O}(10^{-10})$ even for $\phi_0 = 1$. Preheating with such
small coupling is not efficient \cite{Kofman:1997yn}, i.e. reheating
has to proceed via perturbative inflaton decay as we analyzed in
Sec.~\ref{reheat}.

We finally note that the upper bounds on the inflaton couplings we
derived in this Section imply that the rate for inflaton annihilation
reactions, $\phi \phi \rightarrow \chi \bar \chi$ or
$\phi \phi \rightarrow \phi' \bar \phi'$, is always much smaller than
the Hubble rate $H$.  The annihilation rate is given by
$\langle \sigma v \rangle n_\phi$, where $\sigma$ is the relevant
annihilation cross section, $\langle \dots \rangle$ denotes averaging
over the ensemble of inflaton particles, and $n_\phi$ is the inflaton
density. Right after inflation one can estimate
$n_\phi \sim V_{\rm inf} / m_\phi$ and $H \sim H_{\rm inf} \sim \sqrt{V_{\rm inf}}$, but even
at this high inflaton density the annihilation rate is many orders of
magnitude smaller than the Hubble rate. The ratio becomes even smaller
at later times, since $H \propto 1/t$ while $n_\phi \propto 1/t^2$
during matter domination. Therefore inflaton annihilation plays no
role in the dynamics of reheating.

\section{The Scales of Inflation}
\label{scale}

Having derived a lower bound on $\phi_0$ we can discuss the range
of energy scales during and just after inflation that can be
realized in our model. With this we mean both the vacuum energy
during inflation (or, equivalently, the Hubble parameter), and the
range of reheating temperatures after inflation.

Since $\phi_{\rm CMB}$ is very close to $\phi_0$, the inflationary
scale $H_{\rm inf}$ is essentially equal to that at the
inflection-point $\phi_0$. From eqs.~(\ref{inflaton_potential2}) and
(\ref{pa3}) we have
\be \label{energy_scale}
V(\phi_0) = \frac{1}{3} d \phi_0^4 \simeq 2.2 \cdot 10^{-16} \phi_0^6 \,,
\ee
where we have neglected $\beta$ and used $A=-8\phi_0/3$. This
corresponds to a Hubble parameter
\be \label{saddle_scale}
H_{\rm inf} =\sqrt{ \frac{V(\phi_0) }{3}}
\simeq 8.6 \cdot 10^{-9} \phi_0^3\,.
\ee
In the previous section we saw that $\phi_0 \gtrsim 3 \cdot 10^{-5}$;
the lower bound on the bound on the Hubble parameter during inflation is
thus
\be
H_{\rm inf} \gtrsim 2.3 \times 10^{-22} \simeq 0.6 \ {\rm MeV} \,.
\ee
In such a low scale inflationary scenario, the cosmological moduli
problem can be relaxed \cite{Coughlan:1983ci}. Besides the
isocurvature bound of QCD axion can be easily satisfied, making our
model a good candidate to embed QCD axion as dark matter, which can
even allow a wider cosmological window with larger decay constant
$f_a$ \cite{Guth:2018hsa,Ho:2019ayl}. On the other hand, for
$\phi_0 \simeq 1$, $H_{\rm inf}\sim 10^{10}$ GeV is possible, which
allows for the non--thermal production of various particles, and hence
non--standard post--inflationary cosmologies.

It is instructive to compare the inflationary Hubble parameter
(\ref{saddle_scale}) with the change of the inflaton field during one
Hubble time due to the slow--roll of the field. The latter is given by
\be \label{delphi}
\Delta \phi = \frac{|\dot{\phi}|} {H} = \frac{|V'|} {3 H^2}
= \frac{|V'|} {V} = \frac{24 \beta} {\phi_0} = 2.3 \cdot 10^{-5} \phi_0^3\,,
\ee
which is much larger than $H_{\rm inf}/(2\pi)$. This means that even
near the inflection point the dynamics of the inflaton field is
entirely dominated by the classical (SR) equation of motion.

\begin{figure}[ht!]
	\centering
	\includegraphics[width=.6\paperwidth, keepaspectratio]{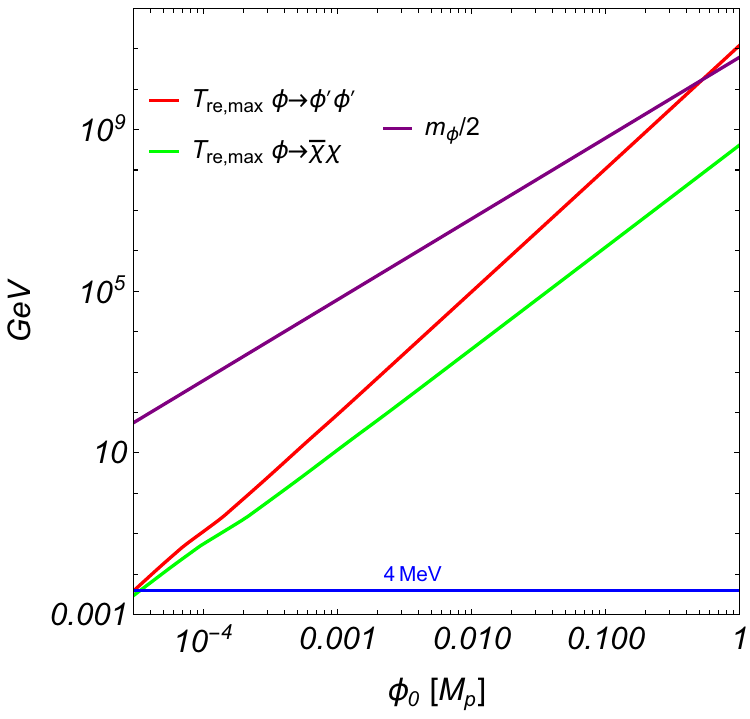}
	\caption {Allowed range of the post--inflationary reheat temperature
		as a function of $\phi_0$ (in the range $[3\cdot10^{-5},1]$), for bosonic (red) and fermionic (green)
		inflaton decays. The blue line shows the lower bound of 4 MeV from
		BBN considerations, and the purple line denotes half of inflaton mass within the parameter space.}
	\label{fig:tre}
\end{figure}

The other energy scale of interest in inflationary model building is
the reheating temperature. As long as we don't fix the relevant
coupling $y$ or $g$, we cannot make a firm prediction; however, the
upper bounds on these couplings that we derived in the previous
Section allow to derive an upper bound on $\Tre$ for given
$\phi_0$. This is shown in Fig.~\ref{fig:tre}, where we have again
used the instantaneous reheating approximation. We see that for
fermionic (bosonic) inflaton decay, the reheating temperatures as high as
$4 \cdot 10^8$ GeV ($10^{11}$ GeV) are possible. This allows for
standard thermal leptogenesis \cite{Davidson:2002qv}. Of course,
the fermionic decay product $\chi$ might itself be right--handed
neutrinos (which contribute $g_\chi = 2$ for each generation), allowing
for non--thermal leptogenesis if the coupling $y$ is (well) below its
upper bound.

The slopes of the curves can be understood as follows. For fermionic
decays, $\Gamma_\phi \propto y^2 m_\phi$, with
$m_\phi \propto \phi_0^2$ from eqs.(\ref{mphi}) and (\ref{pa3}) while
$y_{\rm max}^2 \propto \phi_0^3$ (up to logarithmic corrections) from
the constraint (\ref{ybound}), hence
$T_{\rm re,max} \propto \Gamma_{\phi,{\rm max}}^{1/2} \propto
\phi_0^{5/2}$.  For bosonic decays, $\Gamma_\phi \propto g^2/m_\phi$
and $g_{\rm max}^2 \propto \phi_0^8$ (again up to logarithmic
corrections) from (\ref{gbound}), hence
$T_{\rm re,max} \propto \phi_0^3$. In these simple estimates we have
ignored the dependence of $g_*$ on $\Tre$, which has been included in
Fig.~\ref{fig:tre}. When the temperature is around $0.1$ GeV, the QCD
deconfinement transition happens, leading to a rapid change of
$g_{\star}$ \cite{Drees:2015exa}; this is the reason for the features
in the red and green curves at $T_{\rm re,max}\sim 0.1$ GeV.

Recall from eq.(\ref{mphi}) that
$m_\phi/2 = \sqrt{d} \phi_0 = 6.2 \cdot 10^{10} \ {\rm GeV} \times
(\phi_0/M_{\rm Pl})^2$, which is somewhat above the maximal reheat
temperature for fermionic inflaton decays as shown in
Fig.~\ref{fig:tre}. For fermionic decays a scenario with
$\Tre > m_\phi/2$ is difficult to realize; instead, Pauli blocking
would delay inflaton decays such that $\Tre \lesssim m_\phi/2$. For
bosonic decays $\Tre > m_\phi/2$ is possible, since several relative
soft bosons can combine into a smaller number of more energetic
bosons.

We also remind the reader that the highest temperature of the thermal
background can be considerably higher than $\Tre$
\cite{Giudice:2000ex}; parametrically, in Planckian units
$T_{\rm max} \sim \sqrt{\Tre} H_{\rm inf}^{1/4} \sim \phi_0^{3/4}
\sqrt{\Tre}$. In our case this is indeed always several orders of
magnitude above $\Tre$, with $T_{\rm max} / T_{\rm re,max}$ scaling
like $\phi_0^{-1/2}  (\phi_0^{-3/4})$ for fermionic (bosonic)
inflaton decays. However, for fermionic inflaton decays one also has
to require $T_{\rm max} \leq m_\phi/2$, as we argued above.

\section{Prehistory}
\label{prehis}

So far our analysis has only been concerned with field values
$\phi \lesssim \phi_0$, which we limited to be not larger than $1$ (in
Planckian units). In that sense our model is a ``small field'' model
of inflation.

In this Section we nevertheless wish to briefly describe the dynamics
at much larger field values. After all, except for possible quantum
gravity effects our model can be UV complete, i.e. it might describe
the dynamics also at much larger field values.\footnote{It has been
  conjectured that complete models of quantum gravity ``always''
  contain many relatively light degrees of freedom if the inflaton
  field moves over trans--Planckian field ranges, which means that one
  might lose control over the theory \cite{ArkaniHamed:2006dz,
    Conlon:2012tz}. However, explicit counter--examples in the
  framework of string theory seem to exist \cite{Hebecker:2015tzo}. We
  also note that additional light fields need not affect the dynamics
  of the inflaton, even if they ``generically'' do.}

For field values $\phi \gg \phi_0$ the potential
(\ref{inflaton_potential2}) is dominated by the quartic term
$d \phi^4$. The dynamics in this range is therefore that of quartic
chaotic inflation \cite{Linde:1983gd}. In particular, the deterministic
change of $\phi$ during one Hubble time, $|\dot{\phi}| / H$, will
be smaller than the random variation $H/(2\pi)$ if
\be \label{chaos}
\phi > \phi_{\rm ch, \,min} = 1.2 \cdot 10^3 \phi_0^{-1/3}\,,
\ee
where we have again used eq.(\ref{pa3}) for the strength of the
quartic coupling. If $\phi$ ever satisfied this bound, a period of
``eternal'' inflation started; in fact, in this case it should
continue even now in ``most'' of space. This epoch of eternal
inflation might allow to sample a ``landscape'' of minima of the
(total) effective potential, which seems to be a feature of
superstring theory \cite{Susskind:2003kw}.

Of course, in our patch of the universe eternal inflation must have
ended at some point. It would have been followed by a long period of
deterministic inflation, since for
$\phi \lesssim \phi_{\rm ch, \, min}$ the SR parameters are still very
small. This first phase of deterministic SR inflation ended at
\be \label{end1}
\phi = \phi_{\rm e,1} \simeq \sqrt{12} + \frac{2}{3} \phi_0\,,
\ee
where we have neglected terms of order $\phi^2_0$. This first phase of
deterministic inflation, where $\phi_{\rm ch,\, min} > \phi > \phi_{\rm e,1}$,
lasted for
\be \label{nd1}
N_{\rm det, 1} \simeq 1.8 \cdot 10^5 \phi_0^{-2/3}
\ee
$e-$folds. It should be noted that any initial field value
$\phi_i > \phi_{\rm e,1}$ would lead to large--field SR inflation;
large field inflation is much less sensitive to initial conditions
than small--field inflation \cite{Clough:2016ymm}. Of course, if our
universe indeed underwent a period of eternal inflation, the question
of initial conditions might be moot \cite{Linde:1986fc}.

For $\phi < \phi_{\rm e,1}$ the field underwent fast roll (or overshooting), until it reached
the vicinity of the near--inflection point $\phi_0$ \cite{Drees:2019xpp}. Here we can use an
expansion as in eq.(\ref{delta}) again, but now $\delta$ is negative,
at least initially. SR inflation then starts again once $|\eta_V| < 1$,
which is true for
\be \label{begin2}
\phi < \phi_{\rm b} \simeq \phi_0 \left( 1 + \frac {\phi_0^2}{24}
- \frac {\phi_0^4}{384} + {\cal O}(\phi_0^6) \right)\,.
\ee
Here we have neglected terms of order $\beta$. Eventually $\phi$
reached the value $\phi_{\rm CMB} = \phi_0 ( 1 - \delta_{\rm CMB})$,
with $\delta_{\rm CMB}$ given by eq.(\ref{pa1}). SR inflation with
$\phi_{\rm b} > \phi > \phi_{\rm CMB}$ gave rise to another
\be \label{nd2}
N_{\rm pre-CMB} \simeq 120
\ee
$e-$folds of inflation, with Hubble parameter given by
eq.(\ref{saddle_scale}).

This second deterministic stage of SR inflation would have been
sufficient to completely dilute any relics from possible earlier
large--field inflationary phases, even before density perturbations on
CMB scales were generated. Therefore the ``pre--history'' sketched in
this Section most likely does not have any direct observational
consequences.

\section{Summary and Conclusions}
\label{sum}

In this paper, we have revisited the renormalizable small field
polynomial inflation model. This model can reproduce cosmological data
only if the potential possesses an ``almost'' inflection point
$\phi_0$, such that $\phi \simeq \phi_0$ during inflation. Expanding
in $\phi_0 - \phi$ allowed us to derive accurate analytical
expressions for all relevant quantities. This includes the number of
$e-$folds of inflation after the pivot scale crossed out of the
horizon, $N_{\rm CMB}$, given in eq.(\ref{ncmb}), as well as the power
spectrum, spectral index, its running, and the tensor--to--scalar
ratio $r$, as shown in eqs.(\ref{ns}-\ref{ratio}).

As usual for small--field models of inflation, $r$ is too small to be
detectable by currently conceivable experiments, i.e. a convincing
detection of gravitational waves of inflationary origin would exclude
our model. A second prediction is a negative running of the spectral
index, given by
$\alpha = -1.43 \times 10^{-3} +5.56 \times 10^{-5} \left(N_{\rm
	CMB}-65 \right)$, which might be detectable in future
\cite{Munoz:2016owz}. Note that this is independent of $\phi_0$, which
is the only free parameter of our model once we have fixed the overall
power of the density perturbations, their spectral index, and
$N_{\rm CMB}$.

A complete model also has to provide for a mechanism to reheat the
universe after inflation ends. Here we considered inflaton decays into
either fermions or bosons via trilinear interactions. For given
$\phi_0$ the corresponding coupling strengths can be bounded from
below by demanding that the reheating temperature is sufficiently high
for successful BBN. On the other hand, we showed that the radiative
stability of the inflaton potential near the inflection point leads to
upper bounds on these couplings, which again depend on $\phi_0$. These
constraints on the parameter space are summarized in Fig.~\ref{ps}. In
particular, radiative stability requires $\phi_0 > 3 \cdot 10^{-5}$ in
Planckian units. Within the allowed parameter  space the Hubble
parameter during inflation ($H_{\rm inf}$) can be as low as $\sim 1$ MeV, which makes our model a good candidate to embed QCD axion as dark matter allowing  wider cosmological window \cite{Guth:2018hsa,Ho:2019ayl}. On the other hand, $H_{\rm inf}$ can also be as high
as $10^{10}$ GeV if $\phi_0 \simeq 1$. In this case the reheat
temperature might be as high as $4 \cdot 10^8 \ (10^{11})$ GeV for
fermionic (bosonic) inflaton decays as shown in Fig.~\ref{fig:tre}. We finally showed that our
potential also allows for a phase of ``eternal''
inflation if the field ever was large enough. While this does not
directly affect any observables, it can address conceptual issues
involving the ``landscape'' of superstring theory and the initial
conditions for inflation.

Of course, if large field values are admissible, one can also consider
scenarios where the near inflection point lies at $\phi_0 > 1$.  Since
$\delta_{\rm CMB} \propto \phi_0^2$ the expansion we used in this
paper will no longer work when $\phi_0$ becomes large. Qualitatively
new features will then become possible, including a sizable value of
$r$ and two distinct epochs of eternal inflation.  We will investigate
the large field version of this model in a future publication.

The least attractive feature of this model is that one has to engineer
$\phi_0$ to ``almost'' be an inflection point; specifically, the
parameter $\beta$, which controls the flatness of the potential around
$\phi_0$, has to be of order $10^{-6} \phi_0^4$, see
eq.(\ref{pa2}). Actually, when written in the form of
eq.(\ref{inflaton_potential2}) the finetuning is not obvious; after
all, $\beta$, while small, is not terribly finely tuned. On the other hand,
the coefficient of the cubic term {\em is} tuned. This conclusion
can be avoided only if $A$ and $\beta$ in eq.(\ref{inflaton_potential2})
can be considered to be independent parameters. This is another
example where conclusions about finetuning depend strongly on what
are considered to be independent parameters. At any rate, our upper
bounds on the relevant couplings imply that the model is at least
technically natural, in the sense that radiative corrections are
under control.

On the other hand, the model we consider is renormalizable, and can
thus serve as the inflationary sector of some well motivated
extensions of the standard model of particle physics; examples are the
$\nu $MSM \cite{Asaka:2005an, Shaposhnikov:2006xi}, or the new minimal
standard model (NMSM) \cite{Davoudiasl:2004be} which can explain
cosmological dark matter, neutrino masses and the baryon asymmetry.
This offers avenues for future research. See Ref.~\cite{Bernal:2021qrl} for a recent study along this direction.

\section*{Acknowledgment}
We thank Nicolas Bernal and Fazlollah Hajkarim for useful discussions.

\bibliographystyle{JHEP}
\bibliography{biblio}
\end{document}